\documentclass[letterpaper,twocolumn,10pt]{article}
\usepackage{usenix,epsfig,endnotes}
\usepackage{graphicx}
\usepackage{caption, subcaption}
\usepackage{color}
\usepackage{algorithm}
\usepackage{algpseudocode}
\usepackage{url}

\usepackage{hyperref}
\hypersetup{
  pdfinfo={
  pdfproducer={},
  Title={},
  Subject={},
  Author={},
  Producer={},
  Creator={},
  }
}

\begin{document}

\date{}

\title{Are FPGAs Suitable for Edge Computing?}
\author{
{\rm Saman Biookaghazadeh}\\
Arizona State University
\and
{\rm Ming Zhao}\\
Arizona State University
\and
{\rm Fengbo Ren}\\
Arizona State University
} 

\maketitle

\thispagestyle{empty}

\subsection*{Abstract}
The rapid growth of Internet-of-things (IoT) and artificial intelligence applications have called forth a new computing paradigm--edge computing. In this paper, we study the suitability of deploying FPGAs for edge computing from the perspectives of throughput sensitivity to workload size, architectural adaptiveness to algorithm characteristics, and energy efficiency. This goal is accomplished by conducting comparison experiments on an Intel Arria 10 GX1150 FPGA and an Nvidia Tesla K40m GPU. The experiment results imply that the key advantages of adopting FPGAs for edge computing over GPUs are three-fold: 1) FPGAs can provide a consistent throughput invariant to the size of application workload, which is critical to aggregating individual service requests from various IoT sensors; (2) FPGAs offer both spatial and temporal parallelism at a fine granularity and a massive scale, which guarantees a consistently high performance for accelerating both high-concurrency and high-dependency algorithms; and (3) FPGAs feature 3--4 times lower power consumption and up to 30.7 times better energy efficiency, offering better thermal stability and lower energy cost per functionality.

\section{Introduction} \label{sec:intro}

The Internet-of-Things (IoT) will connect 50 billion devices and is expected to generate 400 Zetta Bytes of data per year by 2020. Even considering the fast-growing size of the cloud infrastructure, the cloud is projected to fall short by two orders of magnitude to either transfer, store, or process such vast amount of streaming data~\cite{fogcomputing}. Furthermore, the cloud-based solution will not be able to provide timely service for many time-sensitive IoT applications~\cite{chiang2016fog}~\cite{winkler2014security}. Consequently, the consensus in the industry is to expand our computational infrastructure from data centers towards the edge. Over the next decade, a vast number of edge servers will be deployed to the proximity of IoT devices; a paradigm that is now referred to as fog/edge computing.

There are fundamental differences between traditional cloud and the emerging edge infrastructure. The cloud infrastructure is mainly designed for (1) fulfilling time-insensitive applications in a centralized environment; (2) serving interactive requests from end users; and (3) processing batches of static data loaded from memory/storage systems. Differently, the emerging edge infrastructure has distinct characteristics, as it keeps the promise for (1) servicing time-sensitive applications in a geographically distributed fashion; (2) mainly serving requests from IoT devices, and (3) processing streams of data from various input/output (I/O) channels. Existing IoT workloads often arrive with considerable variance in data size and require extensive computation, such as in the applications of artificial intelligence, machine learning, and natural language processing. Also, the service requests from IoT devices are usually latency-sensitive. Therefore, having a predictable latency and throughput performance is critical for edge servers.

Existing edge servers on the market are simply a miniature version of cloud servers (cloudlet) which are primarily structured based on CPUs with tightly coupled co-processors (e.g., GPUs)~\cite{HPEEdgeLineEL4000}~\cite{HPEEdgeLineEL1000}~\cite{HPEGL10IOTGATEWAY}~\cite{CISCOUCSC220}. However, CPUs and GPUs are optimized towards batch processing of memory data and can hardly provide consistent nor predictable performance for processing streaming data coming dynamically from I/O channels. Furthermore, CPUs and GPUs are power hungry and have limited energy efficiency~\cite{fowers2012performance}, creating enormous difficulties for deploying them in energy- or thermal-constrained application scenarios or locations. Therefore, future edge servers call for a new general-purpose computing system stack tailored for processing streaming data from various I/O channels at low power consumption and high energy efficiency.


OpenCL-based field-programmable gate array (FPGA) computing is a promising technology for addressing the aforementioned challenges. FPGAs are highly energy-efficient and adaptive to a variety of workloads. Additionally, the prevalence of high-level synthesis (HLS) has made them more accessible to existing computing infrastructures. In this paper, we study the suitability of deploying FPGAs for edge computing through experiments focusing on the following three perspectives: (1) sensitivity of processing throughput to the workload size of applications, (2) energy-efficiency, and (3) adaptiveness to algorithm concurrency and dependency degrees, which are important to edge workloads as discussed above.

The experiments are conducted on a server node equipped with a Nvidia Tesla K40m GPU and an Intel Fog Reference Design Unit \cite{FogReference} equipped with two Intel Arria 10 GX1150 FPGAs. Experiment results show that (1) FPGAs can deliver a predictable performance invariant to the application workload size, whereas GPUs are sensitive to workload size; (2) FPGAs can provide 2.5--30 times better energy efficiency compared to GPUs; and (3) FPGAs can adapt their hardware architecture to provide consistent throughput across a wide range of conditional or inter/intra-loop dependencies, while the GPU performance can drop by up to 14 times from the low- to high-dependency scenarios. 

The rest of the paper is organized as follows: Section~\ref{sec:background} introduces the background; Section~\ref{sec:methodology} describes the methodology; Section~\ref{sec:experiment} discusses experimental results; and Section~\ref{sec:conclusion} concludes the paper.

\section{Background} \label{sec:background}



An FPGA is a farm of logic, computation, and storage resources that can be reconfigured dynamically to compose either spatial or temporal parallelism at a fine granularity. 
Traditional FPGA design requires hardware description languages, such as VHDL and Verilog, making it out of the reach of application developers. The advent of HLS technology~\cite{gajski2012high} has opened enormous opportunities. Today, one can develop FPGA kernel functions in high-level programming languages (e.g., OpenCL~\cite{opencl}) and deploy the compiled hardware kernels in a run-time environment for real-time computing~\cite{intelsdk}.
Note that OpenCL is a universal C-based programming model that can execute on a variety of computing platforms, including CPUs, GPUs, DSP processors, and FPGAs~\cite{singh2011higher}. The recently-extended support of OpenCL by FPGAs has opened the gate for integrating FPGAs into heterogeneous HPC, cloud, and edge platforms.


Different from widely adopted CPUs and GPUs in the cloud, FPGAs come with several unique features rendering them an excellent candidate for edge computing. \textit{First}, unlike GPUs and CPUs that are optimized for batch processing of memory data, FPGAs are inherently efficient for accelerating streaming applications. A pipelined streaming architecture with data flow control can be easily built on an FPGA to process streams of data and commands from I/O channels and generate output results at a constant throughput with reduced latency. 

\textit{Second}, FPGAs can adapt to any algorithm characteristics due to their hardware flexibility. Different from CPUs and GPUs that can mostly exploit only spatial parallelism, FPGAs can exploit both \textit{spatial} and \textit{temporal} parallelism at a finer granularity and on a massive scale. In spatial parallelism, processing elements (PEs) are replicated in space, while data is being partitioned and distributed to these PEs in parallel. In temporal parallelism, processing tasks that have dependency among each other are mapped onto pipelined PEs in series, while each PE in the pipeline can take data with different timestamps in parallel. FPGAs can construct both types of parallelism using their abundant computing resources and pipeline registers \cite{johnston2004implementing}. This unique feature makes FPGAs suitable for accelerating algorithms with a high degree of both data concurrency and dependency. Therefore, FPGAs keep the promise to efficiently serve a wider range of IoT applications. 

\textit{Third}, FPGAs consume significantly lower power compared to CPUs and GPU~\cite{fowers2012performance} for delivering the same throughput, allowing for improved thermal stability and reduced cooling cost. This merit is critically needed for edge servers, considering their limited form factors.




\begin{figure}
\centering
\resizebox{\columnwidth}{!}{%
\includegraphics{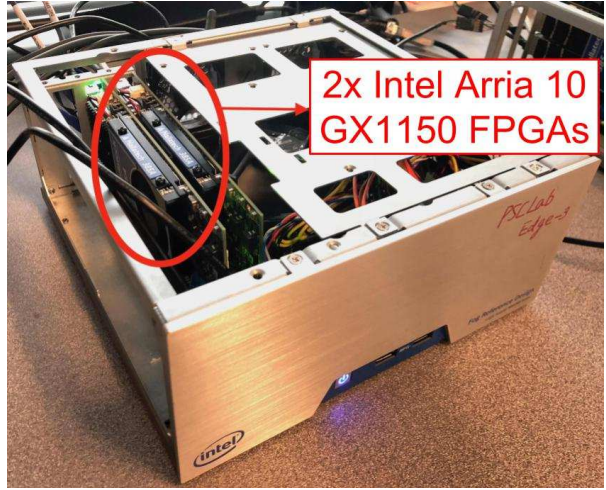}
}
\caption{An Intel Fog Reference Design unit hosting two Nallatech 385A FPGA Acceleration Cards.}
\label{fig:fogref}
\vspace{-1.0em}
\end{figure}

\section{Methodology} \label{sec:methodology}

To confirm and quantify the aforementioned benefits of FPGA-based edge computing, we designed and conducted three sets of experiments to evaluate FPGAs vs. GPUs from the perspectives of (1) \textit{performance sensitivity to workload size}, (2) \textit{adaptiveness to algorithm concurrency and dependency degrees}, and (3) \textit{energy efficiency}.

All the GPU-related experiments were conducted on a server node equipped with a Nvidia Tesla K40m GPU, dual Intel Xeon E5-2637 v4 CPUs, and 64GB of main memory. All the FPGA-related experiments were conducted on an Intel Fog Reference Design unit~\cite{FogReference} (see Figure~\ref{fig:fogref}) equipped with two Nallatech 385A FPGA Acceleration Cards (Intel Arria 10 GX1150 FPGA), an Intel Xeon E5-1275 v5 CPU, and 32GB of main memory. The OpenCL kernels for FPGAs were compiled using Intel FPGA SDK for OpenCL (version 16.0) with Nallatech \textit{p385a\_sch\_ax115} board support packages (BSP). The GPU OpenCL kernels were compiled at runtime using available OpenCL library in CUDA Toolkit 8.0. Results discussed in the next section will show that the FPGA substantially outperforms the GPU in several important aspects, despite that the GPU has a much higher theoretical throughput (4.29TFlops) than the FPGA (1.5TFlops).

\section{Experiment Results} \label{sec:experiment}

\begin{figure}[t!p]
\begin{subfigure}[b]{\columnwidth}
        \centering
        \includegraphics[width=\columnwidth]{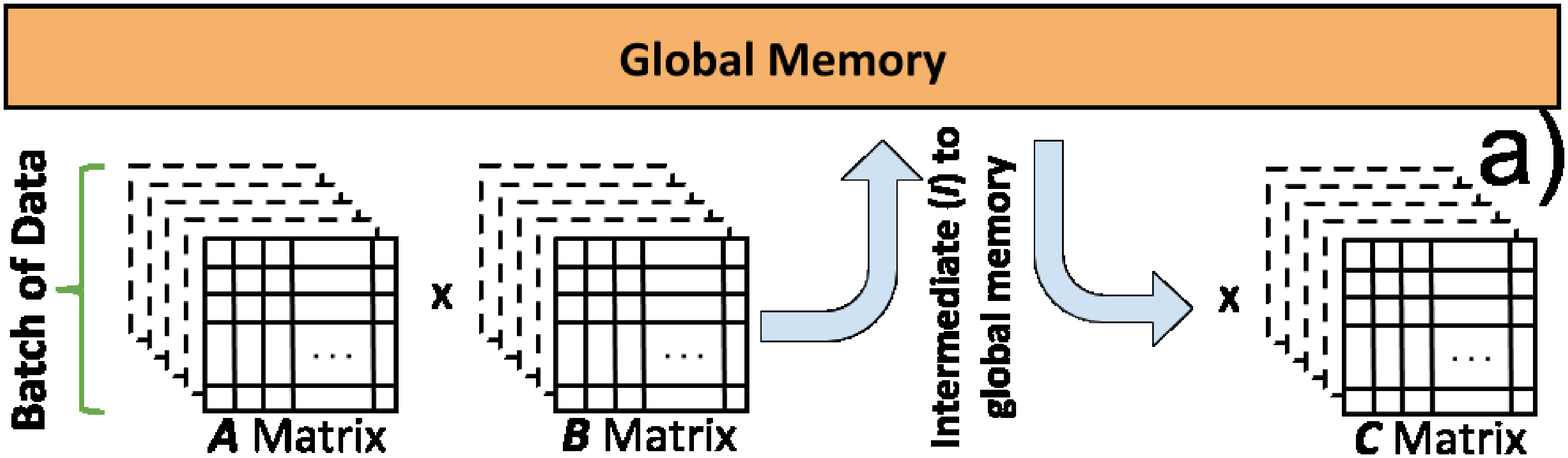}
\end{subfigure}
\begin{subfigure}[b]{\columnwidth}
        \centering
        \includegraphics[width=\columnwidth]{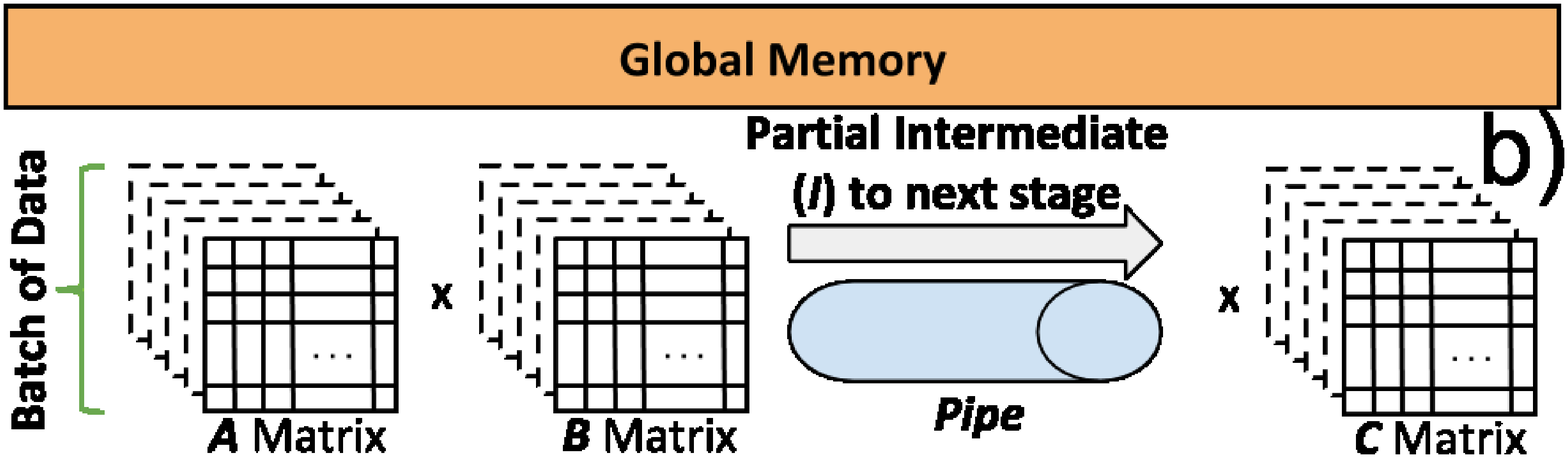}
\end{subfigure}
\caption{Multi-stage matrix multiplication on (a) a GPU and (b) an FPGA.}
\label{fig:MatMul}
\vspace{-1.0em}
\end{figure}

\subsection{Sensitivity to Workload Size} \label{sec:load}

The purpose of this experiment is to demonstrate the sensitivity of FPGA and GPU to workload size. IoT devices are usually latency sensitive and expect predictable latency and throughput from edge servers. We used a two-stage matrix multiplication ($A \times B \times C$) as the benchmark, to model edge workloads. This operation is widely used in linear algebraic algorithms and is generic enough for the purpose of this experiment. All three matrices are of dimension 32x32 and contain single-precision floating-point random numbers. Input matrices are provided as a batch, and the batch size represents the workload size. We varied the batch size between 2 to 2048 in the experiment. The processing throughput (number of matrices/ms) is defined as the ratio of the workload size over the total runtime.

\begin{figure}
\centering
\resizebox{0.8\columnwidth}{!}{%
\includegraphics{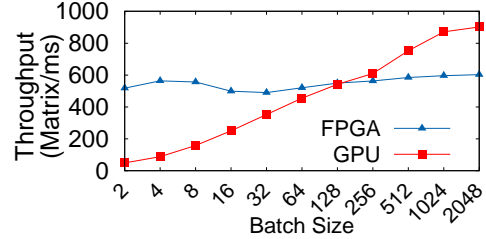}
}
\caption{Sensitivity of matrix multiplication throughput (number of computed matrices per millisecond) sensitivity to batch size (number of matrices received per batch)}
\label{fig:perfcomp}
\vspace{-1.0em}
\end{figure}

Figures~\ref{fig:MatMul}a and~\ref{fig:MatMul}b illustrate the difference of execution flow between the GPU and the FPGA. To exploit spatial parallelism, the GPU must perform $A \times B$ for the entire batch and stores the intermediate results (\textit{I}) in the global memory. Once the writing of \textit{I} is done, the subsequent \textit{I}$\times$\textit{C} can be performed by reading \textit{I} back from the global memory. Differently, the FPGA can also exploit temporal parallelism and utilize dedicated \textit{pipes} (\textit{channels}) to transfer the intermediate results from one stage to another without blocking the execution. The execution of \textit{A}$\times$\textit{B}$\times$\textit{C} is fully pipelined by the streaming architecture implemented in the FPGA, such that the matrix samples can flow in and out of the FPGA through I/O channels one after another without waiting regardless of the batch size.

Figure~\ref{fig:perfcomp} shows the throughput comparison between the GPU and the FPGA across different batch sizes. It is shown that the FPGA can deliver a consistently high throughput by jointly exploiting spatial and temporal parallelism. Specifically, the FPGA outperforms the GPU for small batch sizes (up to 128) in spite of its much lower operating frequency. In contrast, the GPU performance varies largely according to the batch size. GPUs rely on interleaving a large batch of input data to hide the device initialization and data communication overheads. When dealing with small batch size, such overheads will dominate total execution time and degrade the throughput especially when the operations involved have some levels of dependency. 
Overall, the experiment results imply that FPGAs not only are efficient in handling aggregated service requests coming from individual devices in small batch sizes but also can guarantee a consistently high throughput with a well-bounded latency. Therefore, FPGAs are highly suitable for edge computing given the considerable variance in workload size of various IoT applications.

\subsection{Adaptiveness} \label{sec:adapt}

To evaluate how well FPGAs and GPUs adapt to algorithm characteristics, we designed benchmarks to capture two types of dependencies: \textit{data dependency}, which represents the dependency across different iterations of a loop, and \textit{conditional dependency}, which represents the dependency on conditional statements with each iteration of the loop.




Our benchmark resembles an algorithm made of a simple iterative block (\textit{for-loop}) where each iteration performs certain number of operations. The \textit{loop\_length} and \textit{ops} variables define the total number of iterations and the total number of operations per iteration (set to 262144 and 512 in the experiment), respectively. All variables are single-precision in the experiments. Note that the objective of our experiments is to reveal the impact of architecture adaptiveness to algorithm characteristics rather than evaluating the performance for a specific algorithm.

The benchmark captures data dependency by introducing dependency among different iterations of the loop. When there is no data dependency, every single iteration is considered as independent and all the iterations can execute in parallel. With data dependency, the iterations that are dependent on one another need to be executed sequentially as a group. Therefore, by varying the data dependency degree, i.e., the average size of the groups, we can control the data parallelism available in the algorithm using this benchmark. GPU's performance is closely tied to the available data parallelism. In comparison, FPGA can exploit PEs in series and receive iterations regardless of the dependency. Different iterations can co-exist and be executed in the pipeline, while traversing down the connected PEs concurrently.

To introduce conditional dependency, we add \textit{if-else} statements into the iterations of the loop in the benchmark. Half of the iterations are in the \textit{if} block and the other half are in the \textit{else} block. Only the iterations that follow the same branch path can be executed in a data parallel fashion. To reveal the performance impact of conditional dependency, we vary the number of operations in each \textit{if} and \textit{else} block, which affects the initialization overhead and consequently the overall performance. GPU is highly sensitive to conditional dependency, because it can parallelize only the iterations that take the same path at one time. In comparison, FPGA can configure the hardware to include all different execution paths, and use a simple lookup table to direct every thread into the right pipeline and execute all threads at the same time.



In order to get the best performance out of the FPGA and the GPU, the above algorithms were deployed using two different methods. For the GPU, we designed an equivalent OpenCL kernel and deployed it in the NDRange mode to accelerate concurrent operations by exploiting spatial parallelism. For the FPGA, we compiled the FPGA kernel in the \textit{single-threaded} mode to accelerate dependent operations by exploiting temporal parallelism, in which case loop execution is initiated sequentially in a pipelined fashion.

\begin{figure}[t!p]
\begin{subfigure}[b]{0.48\columnwidth}
	\centering
	\includegraphics[width=\columnwidth]{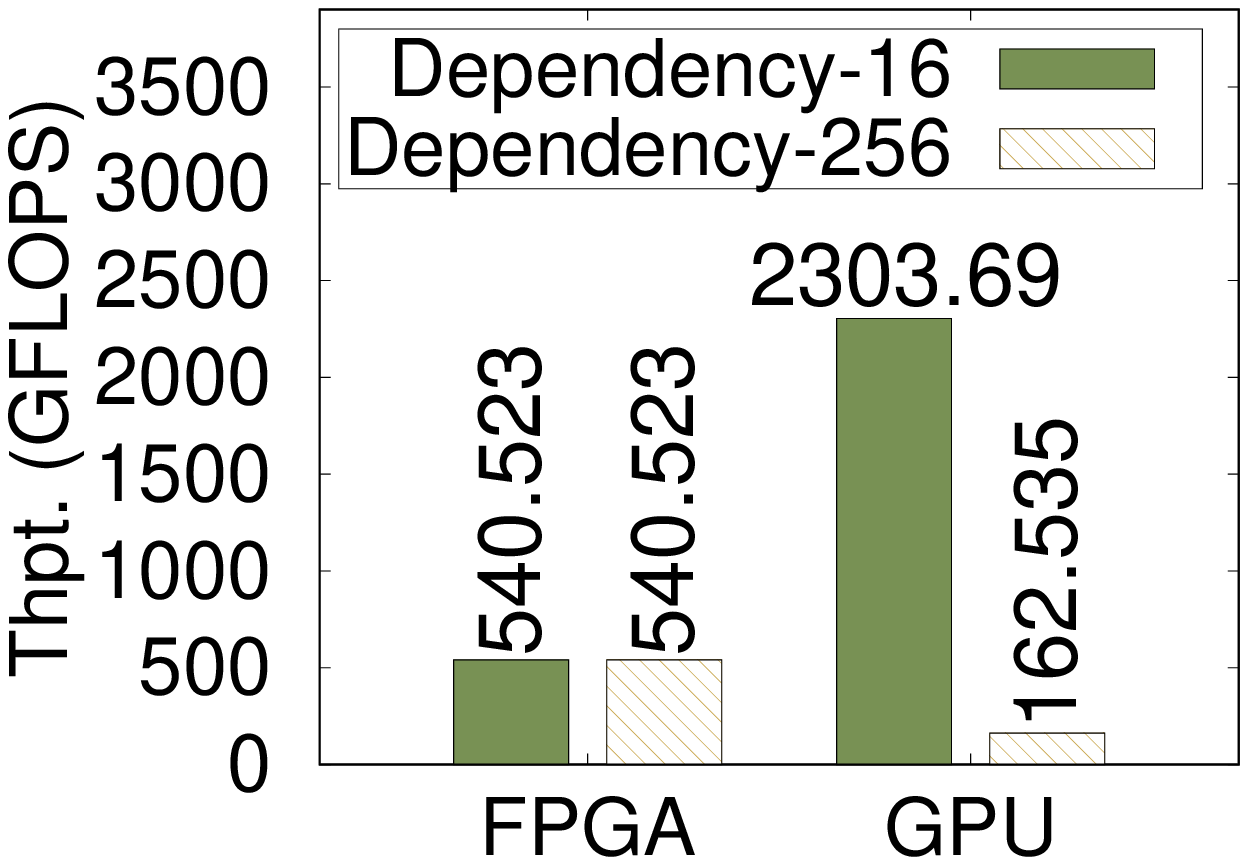}
	\vspace{-0.5em}
	\caption{Raw Throughput}
\end{subfigure}
\begin{subfigure}[b]{0.48\columnwidth}
	\centering
	\includegraphics[width=\columnwidth]{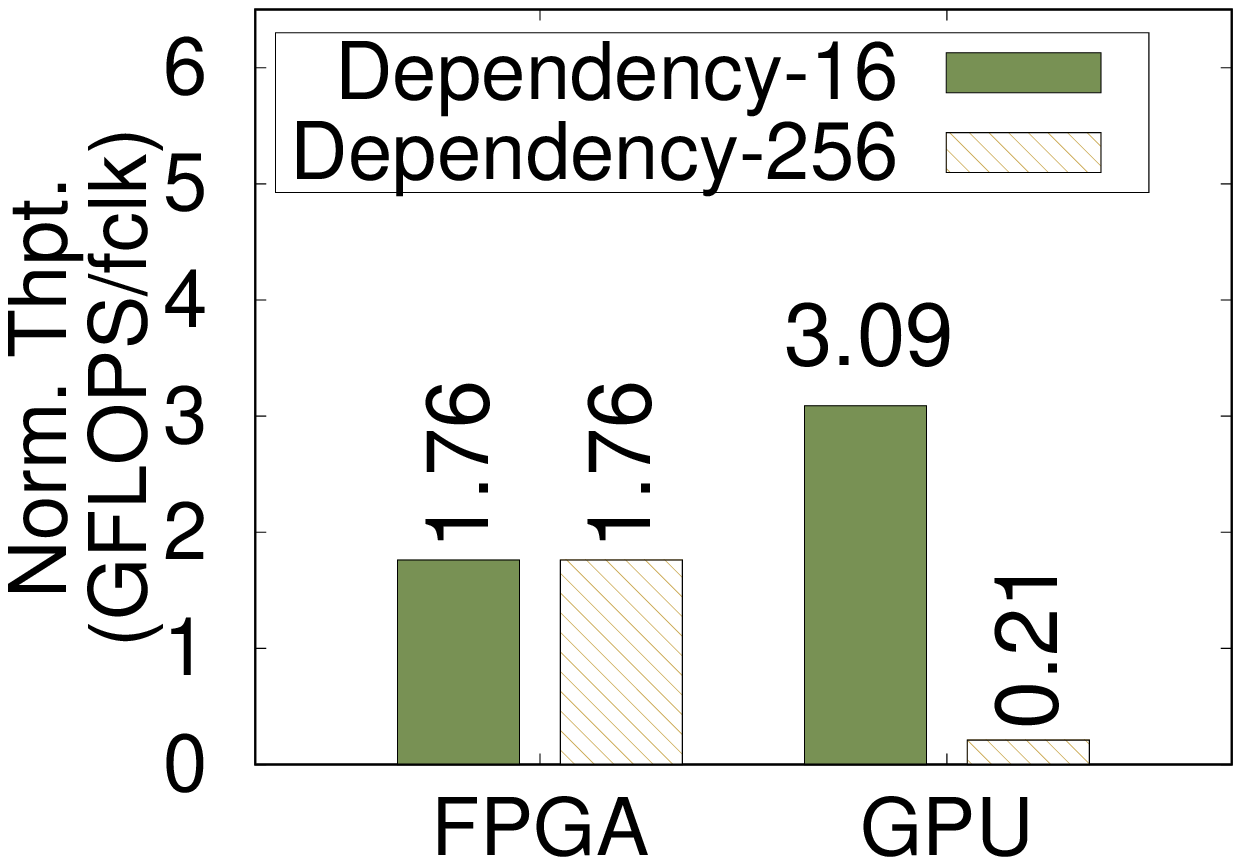}
	\vspace{-0.5em}
	\caption{Normalized Throughput}
\end{subfigure}
\caption{Comparison of (a) raw and (b) normalized throughput at low and high data dependency degrees.}
\label{fig:DepDegree}
\vspace{-1.0em}
\end{figure}

\smallskip
\noindent\textbf{Data Depndency}.
Figures~\ref{fig:DepDegree}a and~\ref{fig:DepDegree}b show the raw and the normalized throughput (to system frequency $f_{clk}$) for both a low (16) and a high (256) data dependecy, respectively. In general, computation throughput is linearly proportional to both $f_{clk}$ and architectural parallelism. The normalized throughput decouples $f_{clk}$  from the evaluation and measures the pure impact of architecture parallelism on throughput. For the GPU, the \textit{base} frequency of the board is used as $f_{clk}$. For the FPGA, $f_{clk}$ is extracted from the full compilation report. It is shown that the GPU performance drops by 14 times from the low to the high data concurrency. As data concurrency increases from 16 and 256, the available data parallelism (the number of loop iterations that can be executed in parallel) for the GPU drops from 16384 to 1024. It is the lack of temporal parallelism that make GPUs hardly adaptive to such changes in concurrency and dependency degrees. On the contrary, the FPGA delivers a stable throughput regardless of such changes. This is because the hardware resources on an FPGA can be reconfigured dynamically to compose either spatial or temporal parallelism (interchangeable) at a fine granularity. As a result, FPGA outperforms GPU by 3.32 folds with the high data concurrency, and this gap is expected to grow as the dependency degree further increases. 

\begin{figure}[t!p]
\centering
\resizebox{0.8\columnwidth}{!}{%
\includegraphics{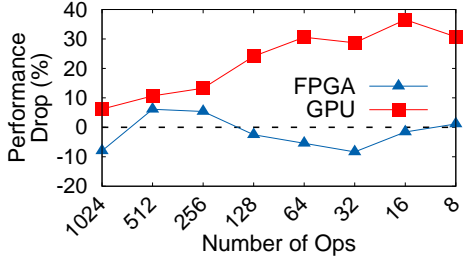}
}
\caption{Performance drop comparison for kernel with conditional statements.}
\label{fig:conditional}
\vspace{-1.0em}
\end{figure}

\smallskip
\noindent\textbf{Conditional Dependency}
Figure~\ref{fig:conditional} shows the performance drop with respect to the conditional dependency introduced by \textit{if-else} statements, as the number of operations \textit{if} and \textit{else} block from 8 to 1024. It shows that the FPGA performance is relatively stable as the conditional dependency increases. For some specific cases, the performance is even increased due to a higher clock frequency compared to the baseline kernel. In contrast, the GPU experiences up to 37.12 times performance drop, compared to baseline kernel with no conditional statements. Branches from the conditional statements cause different threads in a warp to follow different paths, creating instruction replay and resulting in reduced throughput. Figure~\ref{fig:conditional} also shows that having fewer operations in the kernel causes more degradation for the GPU since a smaller kernel requires less computation and incurs relatively higher initialization and data transfer overhead.

\subsection{Energy Efficiency}

To evaluate energy efficiency, we measured the workload throughput divided by its average power usage. To project energy efficiency, the power consumptions of both devices are recorded for all of the experiments. We used the \textit{nvidia-smi} command-line utility and the Nallatech memory-mapped device layer API to query the instant board-level power consumption every 500 milliseconds for the GPU and FPGA, respectively. We then calculated the average power usage by averaging all the power numbers recorded across five trials of each experiment.

\begin{figure}[t!p]
\begin{subfigure}[b]{0.46\columnwidth}
        \centering
        \includegraphics[width=\columnwidth]{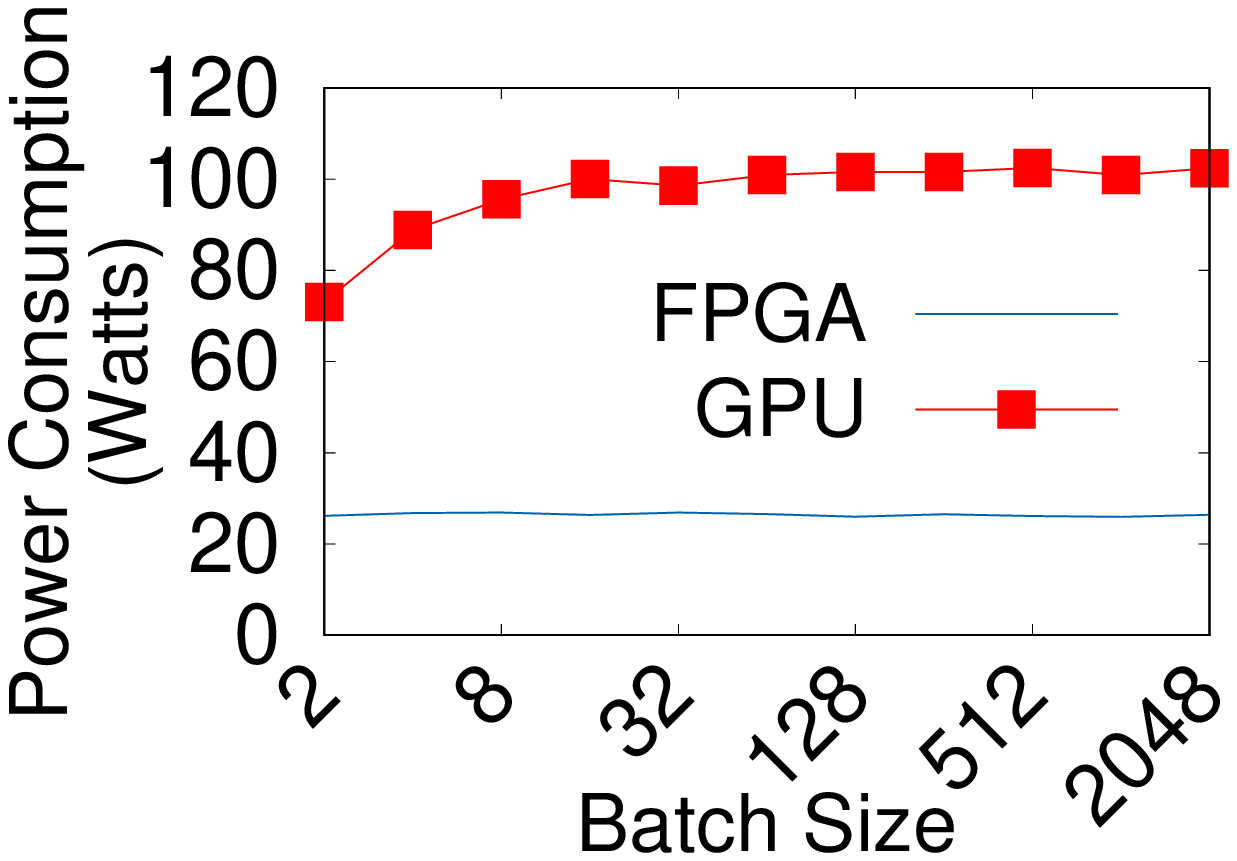}
        \vspace{-1.0em}
        \centering
        \caption{}
        \label{fig:powercomp}
\end{subfigure}
\begin{subfigure}[b]{0.46\columnwidth}
        \centering
        \includegraphics[width=\columnwidth]{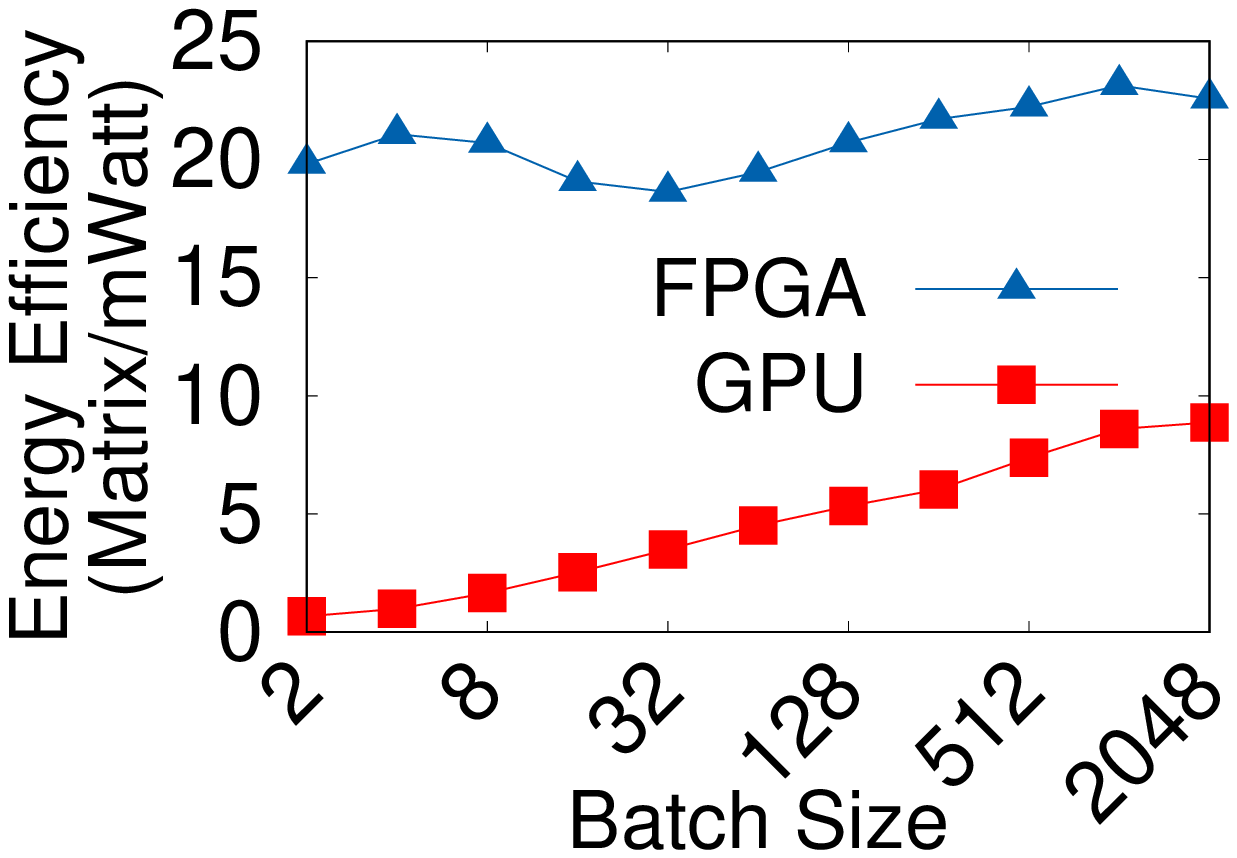}
        \vspace{-1.0em}
        \caption{}
        \label{fig:PowerEff}
\end{subfigure}
\centering
\begin{subfigure}[b]{0.8\columnwidth}
        \centering
        \includegraphics[width=\columnwidth]{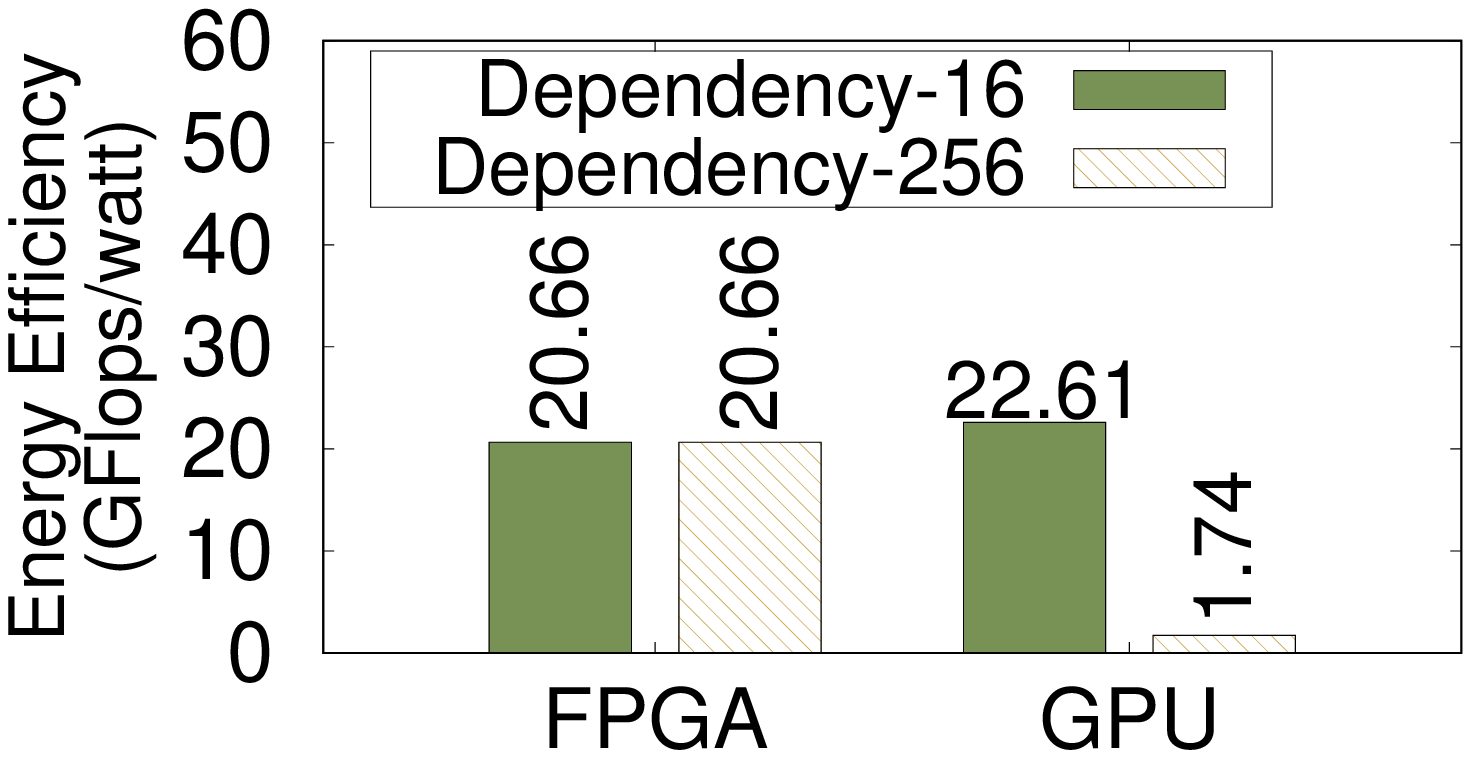}
        \vspace{-1.0em}
        \caption{}
        \label{fig:PowerPseudo}
\end{subfigure}
\vspace{-0.5em}
\caption{The caparisons of (a) power consumption and (b) energy-efficiency for the matrix multiplication tasks and (c) the data dependency benchmark.}
\label{fig:figpower}
\vspace{-1.0em}
\end{figure}

Figure~\ref{fig:figpower}a and ~\ref{fig:figpower}b show the power consumption and energy efficiency comparison for performing the matrix multiplication tasks mentioned in Section~\ref{sec:load}, for different batch sizes. Running at a much lower frequency, the FPGA consistently consumes 2.79--3.92 times lower power than the GPU. Taking into account the performance, it shows that the FPGA can provide 2.6--30.7 times higher energy efficiency than the GPU for executing matrix multiplication. The improvement is prominent, especially for small batch sizes. The low power consumption and the high energy efficiency of the FPGA imply that deploying FPGAs for edge computing can potentially gain better thermal stability at lower cooling cost and reduced energy bill.

Figure~\ref{fig:figpower}c depicts the energy efficiency comparison for running the workloads with different dependency degrees (mentioned in Section~\ref{sec:adapt}). The results show that the FPGA achieves a similar throughput to the GPU for executing the kernels with a high data concurrency degree (low data dependency degree of 16). For the high-data-dependency (degree of 256) workload, the FPGA achieves up to 11.8 times higher energy efficiency than the GPU. Such energy efficiency improvement is expected to further increase as the dependency degree grows. The experiment results indicate that the FPGA is almost on par with the GPU regarding energy efficiency for executing high-concurrency algorithms, while it can significantly outperform the GPU for executing high-dependency algorithms. 

\section{Conclusions and Future work} \label{sec:conclusion}

In this paper, we studied three general requirements of IoT workloads on edge computing architectures and demonstrated the suitability of FPGA accelerators for edge servers. Our results confirm the superiority of FPGAs over GPUs with respect to: (1) providing workload-insensitive throughput; (2) adaptiveness to both spatial and temporal parallelism at fine granularity; and (3) better energy efficiency and thermal stability. Based on our observations, we argue that FPGAs should be considered a replacement or complementary solution for current processors on edge servers.

Based on these results, we will further study FPGA-based edge computing along the following possible directions. First, we aim to extend the study of adaptiveness capabilities of both GPUs and FPGAs by considering other important types of algorithm characteristics. Second, we plan to improve our benchmarking kernels to reflect a wider variety of real-world algorithms. Finally, we will also extend our energy-efficiency study for other types of workloads and algorithm characteristics.

{\footnotesize \bibliographystyle{acm}
\bibliography{edge}}


\end{document}